\documentclass[aps,prl,preprintnumbers,twocolumn,groupedaddress,nofootinbib]{revtex4}

\usepackage[dvips]{graphicx}
\usepackage{color}
\usepackage{amsmath,amssymb,slashed}
\usepackage{hyperref}

\usepackage{changepage}

\usepackage{tikz}
\usetikzlibrary{calc} 
\usetikzlibrary{patterns,snakes} 
\usetikzlibrary{decorations.pathreplacing} 
\usetikzlibrary{decorations.markings} 
\usetikzlibrary{decorations.pathmorphing} 
\usetikzlibrary{positioning}
\usetikzlibrary{arrows.meta}

\newif\ifdraft
\drafttrue
\newif\ifpreprint
\preprinttrue

\def\spa#1.#2{\left\langle#1\,#2\right\rangle}
\def\spb#1.#2{\left[#1\,#2\right]}

\font\tenshuffle=shuffle10 \font\sevenshuffle=shuffle7 \font\fiveshuffle=shuffle7 at 5pt
\def\shuffle{{%
\def\Dshuffle{\mathbin{\hbox{\tenshuffle\char'001}}}%
\def\Sshuffle{\mathbin{\hbox{\sevenshuffle\char'001}}}%
\def\SSshuffle{\mathbin{\hbox{\fiveshuffle\char'001}}}%
\mathchoice{\Dshuffle}{\Dshuffle}{\Sshuffle}{\SSshuffle}}}

\def\beq{\begin{equation}}
\def\eeq{\end{equation}}

\let\Im\relax

\DeclareMathOperator{\Im}{Im}

\newcommand{\eq}{\begin{equation}}
\newcommand{\eqe}{\end{equation}}
\newcommand{\eqa}{\begin{eqnarray}}
\newcommand{\eqae}{\end{eqnarray}}

\newcommand{\p}{\partial}

\newcommand{\ep}{\epsilon}

\newcommand{\bea}{\begin{eqnarray}}
\newcommand{\eea}{\end{eqnarray}}

\newcommand{\bma}{\begin{matrix}}
\newcommand{\ema}{\cr\end{matrix}}

\newcommand{\RR}{\mathbb R}
\newcommand{\CC}{\mathbb C}

\newcommand{\ZZ}{\mathbb Z}

\def\pbw{\p_{\bar w}}
\def\pbx{\p_{\bar x}}
\def\pby{\p_{\bar y}}

\def\bomega{\boldsymbol{\om}}
\def\bkappa{\boldsymbol{\kappa}}

\def\mA{\mathfrak{A}}
\def\mB{\mathfrak{B}}

\def\cG{{\cal G}}

\def\cT{{\cal T}}

\def\bM{{\bf M}}

\def\bz{{\bf z}}

\def\mA{\mathfrak{A}}
\def\mB{\mathfrak{B}}

\def\mJ{\mathfrak{J}}

\def\mL{\mathfrak{L}}
\def\mM{\mathfrak{M}}

\def\mT{\mathfrak{T}}

\def\ZZ{{\mathbb Z}}
\def\RR{{\mathbb R}}

\def\CC{{\mathbb C}}

\def\Im{{\rm Im \,}}

\def\half{{1\over 2}}

\def\p{\partial}

\def\tet{\vartheta}
\def\ep{\varepsilon}
\def\om{\omega}

\def\bomega{\boldsymbol{\om}}
\def\bkappa{\boldsymbol{\kappa}}

\def\sm{\smallskip}
\def\no{\nonumber}

\def\mylength{-0.35cm}
\def\myotherlength{-0.25cm}

\newbox\charbox
\newbox\slabox
\def\s#1{{     
        \setbox\charbox=\hbox{$#1$}
        \setbox\slabox=\hbox{$/$}
        \dimen\charbox=\ht\slabox
        \advance\dimen\charbox by -\dp\slabox
        \advance\dimen\charbox by -\ht\charbox
        \advance\dimen\charbox by \dp\charbox
        \divide\dimen\charbox by 2
        \raise-\dimen\charbox\hbox to \wd\charbox{\hss/\hss}
        \llap{$#1$}
}}

\begin{document}

\preprint{UUITP--21/23}

\title{Cyclic products of higher-genus Szeg\"o kernels, modular tensors and polylogarithms}

\author{Eric D'Hoker$^a$, Martijn Hidding$^b$ and
Oliver Schlotterer$^{b}$}
\affiliation{$^a$ Mani L.\ Bhaumik Institute for Theoretical Physics,  Department of Physics and Astronomy,  University of California, Los Angeles,  CA 90095, USA}
\affiliation{$^b$ Department of Physics and Astronomy, Uppsala University, Box 516, 75120 Uppsala, Sweden}

\begin{abstract}
A wealth of information on multiloop string amplitudes is encoded in fermionic two-point functions known as Szeg\"o kernels. In this paper we show that cyclic products of any number of Szeg\"o kernels on a Riemann surface of arbitrary genus may be decomposed into linear combinations of modular tensors on moduli space that carry all the dependence on the spin structure $\delta$.  The $\delta$-independent coefficients in these combinations carry all the dependence on the marked points and are composed of the integration kernels of higher-genus polylogarithms. We determine the antiholomorphic moduli derivatives of the $\delta$-dependent modular tensors.
\end{abstract}

\maketitle

\section{Introduction}
\vspace{\myotherlength}

In the Ramond-Neveu-Schwarz (RNS) formulation of superstring theory, space-time supersymmetry is implemented via  the Gliozzi-Scherk-Olive (GSO) projection.  On a Riemann surface worldsheet of genus $h$ the GSO projection is realized by summing over the $2^{2h}$ different spin structures of the worldsheet fermions, consistently with modular invariance. At genus one, the Riemann relations between Jacobi $\tet$-functions and properties of modular forms \cite{Mumford, Fay:1973} provide systematic tools for evaluating these spin structure sums explicitly \cite{Tsuchiya:1988va, Stieberger:2002wk, Bianchi:2006nf, Dolan:2007eh, Tsuchiya:2012nf, Broedel:2014vla, Berg:2016wux, Tsuchiya:2017joo}. At higher genus, however, carrying out spin structure sums and thus exposing the simplifications due to space-time supersymmetry presents a significant challenge which is often regarded as a drawback of the RNS formulation for evaluating superstring amplitudes. 

\sm

In a recent paper, the authors made progress towards solving the problem of spin structure summations for the special case of even spin structures at genus two \cite{DHoker:2022xxg}, see also \cite{DHoker:2001jaf,DHoker:2005vch, Tsuchiya:2012nf, Tsuchiya:2017joo, DHoker:2021kks, Tsuchiya:2022lqv} for earlier work on this subject. It was shown in \cite{DHoker:2022xxg} that all the dependence on the spin structure of the cyclic product of an arbitrary number of worldsheet fermion propagators, also known as Szeg\"o kernels, may be reduced to the spin structure dependence of certain modular tensors which are locally holomorphic on Torelli space (the moduli space of Riemann surfaces endowed with a choice of canonical homology basis). 
Thanks to certain \textit{trilinear relations} between these modular tensors, all spin structure dependence was further reduced  to that of the well-known four-point functions. 
The restriction to genus two stems from the fact that the results of \cite{DHoker:2022xxg}, including the existence of the trilinear relations, 
rely heavily on the fact that every genus-two Riemann surface is hyper-elliptic which is generically not the case at higher genus.

\sm

In the present paper, we shall consider cyclic products $C_\delta (\bz) = C_\delta(z_1, \cdots , z_n)$ of $n$ Szeg\"o kernels, 
\bea
C_\delta (z_1, \cdots, z_n) = S_\delta(z_1,z_2) S_\delta (z_2, z_3)  \cdots S_\delta (z_n, z_1)\, ,
\label{defds}
\eea
on a Riemann surface $\Sigma$ of arbitrary genus $h$ and even spin structure $\delta$ (encoding the parity-even part of string amplitudes), and an arbitrary number $n\geq 2$ of points $z_i \in \Sigma$. Generalizations of (\ref{defds}) to open chain products of Szeg\"o kernels may be handled by similar methods and their study is deferred to future work. Throughout, the dependence on the moduli of $\Sigma$ will be suppressed. The Szeg\"o kernel $S_\delta(y,z)$ is a differential $(\half,0)$ form in both $y$ and $z$  which, for even spin structure $\delta$ and generic moduli, obeys the chiral Dirac equation,\footnote{Equation (\ref{Szego}) holds for even spin structure $\delta$ throughout moduli space for genus 2 and at generic moduli for genus $h \geq 3$, in which cases its expression in terms of the prime form and the Riemann $\tet$-functions is given in appendix~A. For odd spin structures at all genera, and for even spin structures at genus $h \geq 3$ on the hyper-elliptic divisor where $\tet[\delta](0)=0$, the presence of Dirac zero modes  requires modifying the right side of (\ref{Szego}) into a projector transverse to the zero modes.  Here, we shall restrict to the case where no zero modes are present, deferring the study of the cases with zero modes to future work.}
\bea
\label{Szego}
\pby S_\delta (y,z) = \pi \delta(y,z) \, .
\eea
As the main result of this paper, we completely disentangle the dependence of $C_\delta(\bz)$ on the points $z_i \in \Sigma$ from the dependence on the spin structure $\delta$ for arbitrary genus $h$ and multiplicity $n$. Specifically, $C_\delta(\bz)$  is  decomposed into the following linear combination
\bea
\label{CF}
C_\delta(\bz) = F^{(0)}(\bz) + \sum_{r=2}^n F^{(r)}_{I_1 \cdots I_r} (\bz) C^{I_1 \cdots I_r}_\delta
\eea
\begin{itemize}
\item[(i)] $C_\delta ^{I_1 \cdots I_r}$ are $\delta$-dependent but $z_i$-independent modular tensors of rank $r$ on Torelli space; 
\item[(ii)]  $F^{(r)}_{I_1 \cdots I_r}(\bz)$ are $\delta$-independent functions that carry all the dependence of $C_\delta(\bz)$ on the points $z_i$. Their  explicit form will be derived here and related to  the construction of higher-genus polylogarithms of \cite{DHoker:2023vax}.
\end{itemize}
Therefore, any spin structure sum over $C_\delta(\bz)$ simplifies to a sum over the
$z_i$-independent modular tensors $C_\delta ^{I_1 \cdots I_r}$.

\sm

We calculate the moduli variations of $C_\delta ^{I_1 \cdots I_r}$, identify components that are locally holomorphic in moduli, and thereby pave the way for their systematic evaluation for arbitrary genus $h \geq 3$  in future work.

\sm 

The functions $F^{(r)}_{I_1 \cdots I_r}(\bz)$ provide the natural mathematical setting in terms of which the integrands of higher-genus superstring amplitudes and their low-energy expansions may be organized. As such, they generalize the Parke-Taylor factors familiar at genus zero \cite{Mafra:2022wml} and the Kronecker-Eisenstein kernels at genus one \cite{Broedel:2014vla}.

\vspace{\mylength}
\section{Abelian differentials and the Arakelov Green function}
\vspace{\myotherlength}

The basic ingredients in our construction are convolutions of Abelian differentials and their complex conjugates as well as the Arakelov Green function to be reviewed below. Let~$\Sigma$ be a compact Riemann surface of genus $h$ without boundary. Its first homology group $H_1(\Sigma, \ZZ)$ supports an intersection pairing $\mJ$ for which we choose a canonical basis of cycles $\mA_I$ and $\mB_J$ with $I,J=1, \cdots, h$ with intersection pairing $\mJ(\mA_I, \mB_J) =\delta _{IJ}$ $= - \mJ(\mB_J, \mA_I)$ and $\mJ(\mA_I, \mA_J)=\mJ(\mB_I, \mB_J)=0$.  A canonical basis of holomorphic Abelian differentials $\bomega_I$ is normalized on $\mA_J$ cycles and provides the periods $\Omega_{IJ}$ on the $\mB_J$ cycles,
\bea
\oint _{\mA_J} \bomega_I =\delta _{IJ} \, ,
\hskip 0.8in
\oint _{\mB_J} \bomega_I =\Omega _{IJ} \, .
\eea
The period matrix $\Omega$ is symmetric $\Omega ^t=\Omega$ while its imaginary part $Y =\Im (\Omega)$ is positive definite. The matrices $Y$ and $Y^{-1}$ with components $Y_{IJ}$ and $Y^{IJ}$, respectively,  may be used to raise and lower indices $I,J$ so that, adopting the Einstein summation convention,  we denote $\bomega ^I = Y^{IJ} \bomega _J$ and $\bar \bomega ^I = Y^{IJ} \bar \bomega_J$. In terms of these differentials, and their expression $\bomega_I= \omega_I(z) dz$ in local complex coordinates $z, \bar z$, we may define a canonically normalized volume form $\bkappa$ on $\Sigma$,
\bea
\bkappa = { i \over 2h} \bomega_I \wedge \bar \bomega^I = \kappa (z) d^2 z\, ,
\hskip 0.6in \int _\Sigma \bkappa =1\, ,
\eea
with coordinate volume form $d^2 z = {i \over 2} dz \wedge d\bar z$.
The Arakelov Green function $\cG(x,y)= \cG(y,x)$ is a single-valued function  $\cG: \Sigma \times \Sigma \to \RR$ uniquely defined by \cite{Falt},\footnote{The Dirac $\delta$-function is normalized by $\int d^2 z \, \delta (z,y) f(z) = f(y)$.} 
\bea
\pbx \p_x \cG(x,y) & = & - \pi \delta (x,y) + \pi \kappa (x) \, ,
\no \\
 \int _\Sigma \kappa (x) \cG(x,y) & = &0\, ,
 \label{defAG}
\eea
whose explicit construction via the prime form $E(x,y)$ may be found in \cite{DHoker:2017pvk}.
Besides its defining equations, $\cG(x,y)$ also satisfies the following useful relations,
\bea
\label{ppG}
\pbx \p_y \cG(x,y) & = & \pi \delta (x,y) - \pi \om_I(x) \bar \om ^I (y)\, ,
 \\
\p_x \p_y \cG(x,y) & = & - \p_x \p_y \ln E(x,y) + \pi \om_I(x) \om^I(y) \, .\no
\eea

\vspace{\mylength}
\section{Modular tensors}
\vspace{\myotherlength}

Linear transformations with integer coefficients that act on $H_1(\Sigma, \ZZ)$ by preserving the intersection pairing~$\mJ$ form the modular group $Sp(2h,\ZZ)$. An element $M \in Sp(2h,\ZZ)$ transforms the homology cycles $\mA_I, \mB_J$ by,
\bea
\left ( \bma \mB \cr \mA \ema \right ) \to M \left ( \bma \mB \cr \mA \ema \right )
\hskip 0.8in 
M = \left ( \bma A & B \cr C & D \ema \right ) \, .
\eea
The modular transformation $M$ acts on the period matrix by $\Omega \to (A \Omega +B) (C \Omega +D)^{-1}$ and on the Abelian differentials by its non-linear $GL(h,\CC)$ representation,
\begin{align}
\label{modab}
\bomega _J & \to  \bomega _{J'} \, R^{J'} {}_J \, , & R & = (C \Omega +D)^{-1}  \, ,
\no \\
\bar \bomega ^I & \to  Q^I {}_{I'} \, \bar \bomega ^{I'} \, , &  Q & = C \Omega +D\, .
\end{align}
Modular tensors $\cT$ of arbitrary rank were defined in \cite{DHoker:2020uid} (see also \cite{Kawazumi:lecture,Kawazumi:paper}) to transform as follows,
\bea
&&
\cT^{I_1 \cdots I_r} {}_{J_1 \cdots J_s} 
\label{tenstrf} \\ && \quad
\to
Q^{I_1}{}_{I_1'} \cdots Q^{I_r}{}_{I_r'} \cT^{I_1' \cdots I_r'} {}_{J_1' \cdots J_s'} R^{J'_1} {}_{J_1} \cdots R^{J'_s} {}_{J_s}\, .
\no
\eea
While the volume form $\bkappa$ and the Arakelov Green function $\cG$ are invariant under the full modular group $Sp(2h,\ZZ)$, the Szeg\"o kernel and its cyclic products  transform via
$S_\delta(x,y) \to S_{\tilde \delta}(x,y)$ and $C_\delta(\bz) \to C_{\tilde \delta}(\bz)$,
where the spin structure $\delta=[\delta',\delta'']$ maps to $\tilde \delta=[\tilde \delta', \tilde \delta'']$ with,
\bea
\left ( \bma
\tilde \delta'' \\ \tilde \delta'
\ema \right )
=
\left ( \bma
A &-B \\ -C&D
\ema \right )
\left ( \bma
\delta'' \\ \delta'
\ema \right )
+\tfrac{1}{2} \,{\rm diag} \!
\left ( \bma
AB^t \\ CD^t
\ema \right )\, .
\label{trfdelta}
\eea
Accordingly, $S_\delta(x,y)$ and $C_\delta(\bz)$ are invariant under the congruence subgroup $\Gamma_h(2) = \{ M \in Sp(2h,\ZZ) \, | \, M \equiv I_{2h \times 2h} \, ({\rm mod} ~ 2) \}$ that preserves each spin structure. The $z_i$-independent building blocks $C_\delta^{I_1\cdots I_r}$ of $C_\delta(\bz)$, to be introduced below, furnish modular tensors of $Sp(2h,\ZZ)$ for which $\delta \to \tilde \delta$ according to (\ref{trfdelta}),
\bea
C^{I_1 \cdots I_r}_\delta
\to
Q^{I_1}{}_{I_1'} \cdots Q^{I_r}{}_{I_r'} C^{I_1' \cdots I_r'}_{\tilde \delta} \, ,
\label{congtrf}
\eea
or modular tensors of $\Gamma_h(2)$ for which $\tilde \delta= \delta$.

\vspace{\mylength}
\section{Descent procedure for $n=2,3$}
\vspace{\myotherlength}

In this section, we shall introduce a simple and systematic procedure by which $C_\delta(\bz)$ may be decomposed into a linear combination of modular tensors $C_\delta ^{I_1 \cdots I_r}$ on Torelli space. As a major simplification of the spin-structure summation in string amplitudes, the coefficients of these tensors containing all the dependence on the points $z_i$ no longer depend on $\delta$. The method is constructive and recursive and will be referred to as the \textit{descent procedure}. 

\sm

The two-point function, namely the case $n=2$  of (\ref{defds}), offers the simplest such relation,\footnote{When no confusion is expected to arise we shall us the abbreviation $z_i \to i$ as arguments of $S_\delta$, $\cG$, and $C_\delta$. The cyclic product $C_\delta (1, \cdots , n)=C_\delta (z_1, \cdots , z_n)$ is a $(1,0)$ form in each point $z_i$ and is  stripped here of its  $dz_1 \wedge \cdots \wedge dz_n$ factor.} 
\bea
\label{1.two}
C_\delta (x,y) & = &  C_\delta ^{IJ} \om_I(x) \om_J(y) + \p_x \p_y \cG(x,y) \, .
\eea
This equation may be deduced by using the Fay identities \cite{Fay:1973} along with the second relation of (\ref{ppG})
resulting in the symmetric modular tensor,
\bea
C_\delta ^{IJ} =  - { \p^I \p^J \tet [\delta](0) \over \tet [\delta] (0)} - \pi Y^{IJ}
\label{rk2c}
\eea
of $\Gamma_h(2)$ with derivatives $\p^J \tet [\delta](0){=} \frac{ \partial }{\partial \zeta_J}\tet[\delta](\zeta) |_{\zeta=0}$ in the Jacobian variety $ \CC^h / (\ZZ^h {+} \Omega \ZZ^h)$ of the genus-$h$ surface \cite{Mumford, Fay:1973}. We note that the tensorial transformation law $C_\delta ^{IJ}  \to Q^I{}_K Q^J{}_L C_{\tilde \delta}^{KL}$ with $Q$ and $\tilde \delta$ given in (\ref{modab}) and (\ref{trfdelta}), respectively, emerges only upon combining the transformations of the two terms on the right side of (\ref{rk2c}).

\sm

The three-point function offers the lowest-order case solved by using the descent equations. We consider the problem at a generic point in moduli space, where the Szeg\"o kernel satisfies equation (\ref{Szego}), so that $C_\delta (1,2,3)$ satisfies the following Cauchy-Riemann equations,  
\bea
\label{1.three.a}
\bar \p _1 C_\delta (1,2,3) & = & \pi \big ( \delta (1,2) - \delta (1,3) \big ) C_\delta(2,3)\, ,
\no \\
\bar \p _2 C_\delta (1,2,3) & = & \pi \big ( \delta (2,3) - \delta (2,1) \big ) C_\delta(1,3)\, ,
\no \\
\bar \p _3 C_\delta (1,2,3) & = & \pi \big ( \delta (3,1) - \delta (3,2) \big ) C_\delta(1,2)\, ,
\eea
with $\bar \partial_{j} =\partial_{\bar z_j}$. The descent proceeds by solving the first equation as a function of $z_1$ with the help of the Arakelov Green function defined by (\ref{defAG}). 
The solution is, however, not unique as the $\bar \p_1$ operator acting on $(1,0)$ differentials has a non-trivial kernel spanned by the holomorphic Abelian differentials of $\Sigma$. Thus, the general solution may be expressed as follows, 
\bea
\label{step1}
C_\delta (1,2,3) & = &  \om_I(1) C_\delta^I  (2,3)  
 \\ &&
- \big ( \p_1 \cG(1,2) - \p_1 \cG(1,3) \big ) C_\delta (2,3)\, ,
\no
\eea
where $C^I_\delta(2,3)$ is independent of $z_1$. In view of the modular invariance of $C_\delta(\bz)$ and $\cG(1,a)$, the coefficient $C^I_\delta(2,3)$ transforms as a modular tensor  of $\Gamma_h(2)$,\, i.e.\ according to 
(\ref{congtrf}) with $r=1$. The descent  proceeds by evaluating the Cauchy-Riemann operators $\bar \p_2$ and $\bar \p_3$ on (\ref{step1}), using the second equation  of (\ref{1.three.a}), the relation $\bar \p_2 C_\delta(2,3) = \pi \p_2 \delta(2,3)$, and the second relation in (\ref{ppG}), and we obtain after some simplifications,  
\begin{align}
\label{1.three.b}
\bar \p_2 C^I_\delta(2,3) & =   \pi  \delta(2,3) C_\delta ^{IJ} \om_J(3) - \pi  \bar \om ^I(2) C_\delta(2,3)  \, ,
\\
\bar \p_3 C^I_\delta(2,3) & =   - \pi  \delta(2,3) C_\delta ^{IJ} \om_J(2) + \pi  \bar \om ^I(3) C_\delta(2,3) \, .
\no \quad
\end{align}
To solve  the first equation of (\ref{1.three.b}) in $z_2$ we first decompose $C_\delta(2,3)$ using (\ref{1.two}),
\begin{align}
\label{pC2}
\bar \p_2 C^I_\delta(2,3)  =  & \, \pi  C_\delta ^{JK} \big ( \delta(2,3) \delta ^I_J - \bar \om^I(2)  \om_J(2) \big ) \om_K(3) 
  \no \\ &
   - \pi  \bar \om ^I(2) \p_2 \p_3 \cG (2,3) \, .
 \end{align}
Both lines of the right side are separately integrable since their respective integrals over $\Sigma$ vanish. The following convolutions involving Abelian differentials and the Arakelov Green function, 
 \begin{align}
 \Phi ^I{}_{\! J}(x) & =  \int _\Sigma d^2 z \, \cG(x,z) \bar \om^I(z) \om_J(z) \, ,
 \no \\
 \cG^I(x,y) & =  \int _\Sigma d^2 z \, \cG(x,z) \bar \om^I(z) \p_z \cG(z,y) \, ,
 \label{convols}
\end{align} 
solve the corresponding differential equations, 
\begin{align}
\pbx \p_x    \Phi ^I{}_{\! J}(x) & =  \pi \, \kappa (x) \delta ^I_J - \pi \, \bar \om^I(x) \om_J(x) \, ,
\no \\
\pbx \p_x \cG^I(x,y) & =  - \pi \, \bar \om^I(x) \p_x \cG(x,y) \, .
\end{align}
More succinctly, the special combination defined by,
\begin{align}
f^I{}_{\!J }(x,y) = \p_x  \Phi ^I{}_{\! J}(x)  - \p_x \cG(x,y) \delta ^I_J \, ,
\end{align}
solves the differential equation,
\begin{align}
\pbx f^I{}_{\! J} (x,y) = \pi \delta ^I_J \delta(x,y) - \pi \bar \om^I(x) \om_J(x) \, ,
\end{align}
so that the general solution to (\ref{pC2}) is given as follows,
\bea
\label{step2}
  C^I_\delta(2,3) & = & \om_J(2) C_\delta^{IJ} (3) + f^I{}_{\! J}(2,3) C_\delta^{JK} \om_K(3) 
  \no \\ &&
  + \p_2 \p_3 \cG^I(2,3) \, .
 \eea
Finally, we determine the modular tensor $C_\delta^{IJ} (3)$ from the $\bar \p_3$ derivative of (\ref{step2})  using the second equation in (\ref{1.three.b}), and the following relations, 
\bea
\pby  f^I{}_{\!J} (x,y) & = & - \pi \delta ^I_J \big ( \delta(x,y) -  \om_K(x) \bar \om^K(y) \big ) \, ,
\no \\
\pby \p_x \cG^I(x,y) & = & -\pi f^I{}_{\! J} (x,y) \bar \om^J(y) \, .
\eea 
We obtain an integrable differential equation, 
\beq
\bar \p_3 C_\delta^{IJ} (3)  = 
 \pi \big( \bar \om^I(3) C^{JK}_\delta
-  \bar \om ^J(3) C^{IK}_\delta \big) \om_K(3)  \, ,
\label{antiholo3}
\eeq
whose integral may be obtained in terms of $\Phi$ in (\ref{convols}), 
\bea
\label{step3}
C_\delta ^{IJ} (3) & = &
\om_K(3) C_\delta^{IJK} 
 \\ &&
- C_\delta^{JK} \p_3  \Phi ^I{}_{\! K}(3)   + C_\delta^{IK} \p_3  \Phi ^J{}_{\! K}(3)  \, ,
\no
\eea
where $C_\delta^{IJK}$ is a $z_i$-independent  modular tensor. The relations (\ref{step1}), (\ref{step2}) and (\ref{step3})  give a formula for $C_\delta ^{IJK}$,
\bea
\label{int.three}
C_\delta ^{I_1 I_2 I_3} & = & \bigg( \prod_{i=1}^3 
\int _\Sigma d^2 z_i \, \bar \om ^{I_i} (z_i) \bigg)  C_\delta (z_1, z_2, z_3)
\eea
which proves that $C^{IJK}_\delta$ inherits the total antisymmetry in $I,J,K$ from total antisymmetry of  $C_\delta(x,y,z)$ in $x,y,z$, 
\bea
C^{IJK}_\delta = C^{[IJK]}_\delta \, .
\label{3antisymm}
\eea
Eliminating $C^I_\delta(2,3)$ and $C^{IJ}_\delta (3)$  from (\ref{step1}), (\ref{step2}) and (\ref{step3})
expresses $C_\delta(1,2,3)$ in the general form (\ref{CF}) where
\bea
F^{(0)}(\bz) & = &  - \big ( \p_1 \cG(1,2) - \p_1 \cG(1,3) \big )  \p_2 \p_3 \cG(2,3)  
\no \\ &&
  + \om_I(1) \p_2 \p_3 \cG^I(2,3) \, ,
\no \\
F^{(2)}_{JK} (\bz) & = & ( \om_J(1) \om_I(2) - \om_I(1) \om_J(2) ) \p_3 \Phi ^I{}_{\! K}(3) 
\no \\ && 
 - \big ( \p_1 \cG(1,2) - \p_1 \cG(1,3) \big )  \om_J(2) \om_K(3)
 \no \\ &&
 + \om_I(1) f^I{}_{\! J} (2,3) \omega_K(3)\, ,
 \no \\ 
 F^{(3)}_{IJK} (\bz) & = & \om_I(1) \om_J(2) \om_K(3) \, .
 \label{fat3pt}
\eea
Clearly, all spin-structure dependence has been reduced to the modular tensors $C_\delta^{IJK}$ and $C_\delta^{IJ}$, while all the dependence on the points $z_1, z_2, z_3$  enters via the $\delta$-independent single-valued functions $\Phi, {\cal G}$ and~$f$.

\vspace{\mylength}
\section{Convolutions and modular tensors}
\label{sec:modten}
\vspace{\myotherlength}

The descent procedure for higher $n$ necessitates higher-rank generalizations of the tensors $\cG^I(x,y)$ and $\Phi^I{}_J(x)$ in (\ref{convols}) obtained from the following convolutions of Arakelov Green functions and Abelian differentials with $r \geq 2$,
\begin{align}
\label{1.C.1}
\!\!\Phi ^{I_1 \cdots I_r} {}_{\! J} (x) & =  
\int _\Sigma d^2 z \, \cG(x,z) \, \bar \om ^{I_1} (z) \, \p_z \Phi ^{I_2 \cdots I_r} {}_{\!J} (z) \, ,
 \\
\!\! \cG^{I_1 \cdots I_r} (x,y) & =  \int _\Sigma d^2 z \, \cG(x,z) \, \bar \om ^{I_1} (z) \, \p_z \cG^{I_2 \cdots I_r} (z,y)  \, ,
\no
\end{align}
which frequently occur in the combination,
\beq
\label{1.C.3}
\!\!\!f^{I_1 \cdots I_r} {}_{\!J} (x{,}y)  = \p_x \Phi ^{I_1 \cdots I_r} {}_{\! J} (x)  
- \p_x \cG^{I_1 \cdots I_{r-1}}(x{,}y) \, \delta ^{I_r} _J \, .
\eeq

\noindent
Conversely, $\p \cG$ and $\p \Phi$ may be obtained as the trace and traceless parts of $f$. The functions $\Phi, \cG$ and $f$ transform as modular tensors  under $Sp(2h,\ZZ)$ and furnish the integration kernels in the recent construction of higher-genus polylogarithms \cite{DHoker:2023vax} (see also \cite{Enriquez:2011,Enriquez:2021, Enriquez:2022} for different approaches to higher-genus polylogarithms in the mathematics literature). More specifically, the higher-genus polylogarithms in \cite{DHoker:2023vax} are defined though iterated integrals over a flat connection whose entire dependence on marked points on $\Sigma$ is expressible in terms of $\omega_I(x)$ and the tensor functions (\ref{1.C.3}). One may recast the convolutions used to define $\Phi, \cG$ and $f$ given in (\ref{1.C.1}) in terms of recursive differential equations obtained for $r \geq 2$ from the trace or traceless part of,
\begin{align}
\pbx f^{I_1 \cdots I_r}  {}_{\!J} (x,y) & =  - \pi \bar \om^{I_1} (x) f^{I_2 \cdots I_r} {}_{\!J} (x,y)  \, ,
\label{1.C.6} \\
\pby f^{I_1 \cdots I_r} {}_{\!J} (x,y) & =   \pi \, \delta ^{I_r} _J \, f^{I_1 \cdots I_{r-1}} {}_{\!K} (x,y) \bar \om ^K(y)  \, .\notag
\end{align}
Based on the tensor functions in this section, the descent procedure introduced  for $n\leq 3$ may be 
extended to arbitrary $n$, see appendix B for the case $n=4$.

\vspace{\mylength}
\section{Descent procedure for any $n$}
\vspace{\myotherlength}

In this section, we apply the descent procedure to the case of arbitrary $n \geq 3$. Inspection of the results (\ref{fat3pt}) 
for the special case $n=3$ (and appendix B for $n=4$) shows that both the differential relations between the various intermediate tensor-functions $C_\delta^{I_1 \cdots I_j}(j{+}1, \cdots, n)$ and the recursive decomposition of $C_\delta(1,\cdots,n)$ expose a simple and important pattern that may be extended and proven for arbitrary~$n$. While the differential relations can be found in appendix C, the integrated relations for $j=1, \cdots, n{-}2$ are given by,
\begin{align}
&C_\delta ^{I_1 \cdots I_{j-1} } (j, \cdots , n) 
 = 
 \om _J (j) C_\delta ^{I_1 \cdots I_{j-1} J} (j{+}1, \cdots, n) 
 \label{alln.00} \\
 &\quad \! \! \! \! \! \! \! 
+ \sum_{i=1}^{j-1} f^{I_{j-1} I_{j-2} \cdots I_i}{}_{\!J}(j, j{+}1) C_\delta ^{I_1 \cdots I_{i-1} J} (j{+}1, \cdots , n)  
\notag \\ 
&\quad \! \! \!  \! \! \! \! 
 - \p_j \big (   \cG^{I_{j-1} \cdots  I_1} (j,j{+}1) {-}   \cG^{I_{j-1} \cdots  I_1} (j,n) \big )
C_\delta (j{+}1,\cdots, n)    \, ,
\notag 
\end{align}
which successively express the dependence of $C_\delta(1,\cdots,n)$ on $z_j$
in terms of $\omega_J(j)$ and the convolutions of the previous section.
The cases $j=n$ and $j=n{-}1$ require separate formulas,
\begin{align}
&C_\delta^{I_1 \cdots I_{n-2} }(n{-}1,n) 
= \omega_J(n{-}1) C_\delta^{I_1\cdots I_{n-2} J}(n)
\label{alln.01}\\
 &\quad
+\sum_{i=2}^{n-2} 
 f^{I_{n-2} I_{n-3}\cdots I_i }{}_{\!J}(n{-}1,n) C_\delta^{ I_{1} I_2\cdots I_{i-1} J}(n)
 \notag \\
&\quad +  f^{I_{n-2} I_{n-3}\cdots I_1 }{}_{\!J}(n{-}1,n) C_\delta^{JK} \omega_K(n)
\notag \\
&\quad + \partial_{n-1} \partial_n {\cal G}^{I_{n-2}\cdots I_2 I_1}(n{-}1,n) 
\notag
\end{align}
as well as
\begin{align}
&C_\delta^{I_1 \cdots I_{n-1} }(n) = 
 \omega_J(n) C_\delta^{I_1\cdots I_{n-1}J}
+ \! \! \! \! \! \! \sum_{1\leq i\leq j \atop{(i,j) \neq (1,n-1)}}^{n-1} \! \! \! \!  (-1)^{i-1} 
\label{alln.02} \\
&\quad \times
 \p_n \Phi ^{ I_1 I_2 \cdots I_{i-1} \shuffle I_{n-1} I_{n-2} \cdots  I_{j+1}  }{}_{\!M}(n) C_\delta^{I_i I_{i+1}\cdots I_j M}\, , \notag
\end{align}
where the shuffle of multi-indices $\vec{I}$, $\vec{J}$ is understood as,
\bea
\Phi^{\vec{I} \shuffle \vec{J} }{}_{\!M}(n) = \sum_{\vec{K} \in \vec{I} \shuffle \vec{J} }
\Phi^{\vec{K} }{}_{\!M}(n) \, .
\eea
Combining (\ref{alln.00}) to (\ref{alln.02}) determines the $z_i$-dependent
constituents $F^{(r)}_{I_1 \cdots I_r}(\bz)$ of $C_\delta(\bz) $ in the notation of (\ref{CF}).

We conclude this section with integral representations of the modular tensors $C^{I_1 \cdots I_n}_\delta$ in terms of $C_\delta(1,\cdots, n)$,
\bea
\label{int.n}
C_\delta ^{I_1 \cdots I_n}  = \bigg( \prod_{i=1}^n 
\int _\Sigma d^2 z_i \, \bar \om ^{I_i} (z_i) \bigg)  C_\delta (z_1, \cdots, z_n) \, ,
\eea
which generalize (\ref{int.three}) to arbitrary $n$ and imply 
dihedral (anti)symmetry $C_\delta ^{I_1  I_2 \cdots I_n} =C_\delta ^{I_2 \cdots I_n I_1}
=(-1)^n C_\delta ^{ I_n \cdots I_2 I_1}$.

\vspace{\mylength}
\section{Variations in moduli}
\vspace{\myotherlength}

While the modular tensor $C^{I_1 \cdots I_n}_\delta $ enters a term in the descent of $C_\delta(z_1, \cdots, z_n)$ that is meromorphic in the points $z_i \in \Sigma$, it is not, in general, meromorphic in moduli. We will study anti-holomorphic derivatives in moduli through complex structure variations $\delta_{\bar w \bar w}$ which, in a conformal field theory setting, amount to an insertion of the stress tensor at a point $w \in \Sigma$ \cite{Verlinde:1986kw, Verlinde:1987sd, DHoker:1988pdl,DHoker:2014oxd}. The anti-holomorphic $\delta_{\bar w \bar w}$-variation of $C_\delta^{I_1 \cdots I_n}$ may be  constructed via the relations $\delta_{\bar w \bar w} \Omega_{IJ}= \delta_{\bar w \bar w} \om_I(z)= \delta_{\bar w \bar w}S_\delta(x,y)=0$,
\bea
\label{var1}
\delta _{\bar w \bar w} \, \bar \Omega _{IJ} & = & - 2 \pi i \, \bar \om_I (w) \, \bar \om_J(w) \, ,
\no \\
\delta _{\bar w \bar w} \, \bar \om^I(x) & = & - \bar \om ^I(w) \, \pbx \pbw \cG(x,w) \, ,
\eea
so that we have $ \delta_{\bar w \bar w}C_\delta(\bz)=0$.  While the variation of the two-point function $\delta_{\bar w \bar w}  C_\delta^{IJ} = \pi^2  \bar \omega^I(w) \bar \omega^J(w)$ readily follows from (\ref{rk2c}), the variation $ \delta_{\bar w \bar w} C_\delta ^{I_1 \cdots I_n}$ at $n\geq 3$ may be derived from the integral representation (\ref{int.n}),
\begin{align}
\label{var2}
\delta _{\bar w \bar w} C_\delta ^{I_1 \cdots I_n}
= & \, \pi \bar \om^{I_1} (w) \pbw \Big ( F_\delta ^{I_2  | I_3 \cdots I_n} (w)
 \\ & 
 ~ - F_\delta ^{I_n | I_2 \cdots I_{n-1} } (w) \Big )  + \hbox{ cycl}(1,\cdots, n)
\no
\end{align}
in terms of $w$-dependent modular tensors $F_\delta$ of $\Gamma_h(2)$,
\bea
F_\delta ^{I_2 | I_3 \cdots I_n} (w)  = 
\int _\Sigma d^2 z_2 \, \cG(w,2) \, \bar \om^{I_2} (2)  C_\delta^{I_3 \cdots I_n}  (2) \, .
\eea
Total symmetrization in $n \geq3$ indices $I_1, \cdots, I_n$ cancels the right side of (\ref{var2}), implying that the left side obeys,
\bea
\delta _{\bar w \bar w} C_\delta ^{(I_1 \cdots I_n)} =0 \, ,
\eea
and the totally symmetric modular tensor $C_\delta ^{(I_1 \cdots I_n)} $ is actually a holomorphic modular tensor on Torelli space.

\vspace{\mylength}
\section{Specialization to genus $\leq 2$}
\vspace{\myotherlength}

At genus one, each tensor $C_\delta^{I_1 I_2\cdots I_n} \rightarrow C_\delta^{11\cdots 1}$ has a single component (which vanishes at odd rank $n$) and is a degree-two polynomial in $e_\delta \in \{ \wp(\tfrac{1}{2}),\wp(\tfrac{\tau}{2}),\wp(\tfrac{1+\tau}{2})\}$ at even $n$ \cite{Tsuchiya:2012nf, Tsuchiya:2017joo}, where we set $\Omega_{11}=\tau$. The Weierstrass $\wp$ function at genus one is evaluated at the half-periods associated with the three even spin structures, and the coefficients of $e_\delta^2,e_\delta^1,e_\delta^0$ are combinations of holomorphic Eisenstein series, see appendix D for examples at $n\leq 8$.

\sm

At genus two, the results of \cite{DHoker:2022xxg}, translated into the language of $\tet$-functions, again organize the entire spin-structure dependence of $C_\delta(1,\cdots, n)$ into degree-two polynomials in the components of the $z_i$-independent, symmetric modular tensor of $\Gamma_{h=2}(2)$,
\bea
\mL_\delta ^{IJ} =  { 2 \pi i \over 10} \p^{IJ} \ln \bigg ( { \Psi_{10}  \over \tet [\delta](0)^{20}} \bigg) \, ,
\label{defelldelta}
\eea
where $\Psi _{10} $ denotes the Igusa cusp form (a Siegel modular form of weight ten), and $\p^{IJ}= \frac{1}{2} (1{+}\delta^{IJ})\frac{\partial}{\partial \Omega_{IJ}}$ are the moduli derivatives. In the two-point function,
\begin{align}
C_\delta (x,y) &= \om_I(x) \om_J(y) \mL_\delta ^{IJ} - \wp (x,y) \, ,
\label{Cph2} \\
\wp (x,y)  &=  \p_x \p_y \ln E(x,y) + { 2 \pi i \over 10}  \p^{IJ} \ln \Psi_{10} \, ,
\notag 
\end{align}
the modular tensor $\mL_\delta ^{IJ}$ in (\ref{defelldelta}) offers an alternative to capturing the $\delta$-dependence through the modular tensor $C_\delta^{IJ}$ in (\ref{rk2c}), and $\wp$ denotes the $Sp(4,\mathbb Z)$-invariant genus-two generalization of the Weierstrass function.

\sm

A first key result of \cite{DHoker:2022xxg} is that all spin structure dependence of $C_\delta(\bz)$ at genus two and any multiplicity $n$ may be reduced to a linear combination of tensor powers of $\mL_\delta^{IJ}$. A second result of \cite{DHoker:2022xxg} is that the tensor $\mL_\delta^{IJ}$ satisfies the \textit{trilinear relations},
eliminating any tensor power higher than two and leading to the major simplification, 
\bea
C^{I_1 \cdots I_n} _\delta & = & (\mT_n^{(2)}) ^{I_1 \cdots I_n}{}  _{J_1 J_2 J_3 J_4 } \, \mL_\delta ^{J_1 J_2} \, \mL_\delta ^{J_3J_4} 
\label{majsimp} \\ &&
+ (\mT_n^{(1)}) ^{I_1 \cdots I_n}{} _{J_1 J_2} \, \mL_\delta ^{J_1 J_2} + (\mT_n^{(0)}) ^{I_1 \cdots I_n}  \, .
\no
\eea
The modular tensors $\mT_n^{(i)}$ of $Sp(4,\mathbb Z)$ are independent of~$z_i$ and $\delta$ but, just as $C^{I_1 \cdots I_n} _\delta $, they are not necessarily locally holomorphic in moduli.
Since spin-structure independent modular tensors of odd rank must vanish at genus two, the modular tensors $\mT_n^{(i)}$ and therefore $C^{I_1 \cdots I_n} _\delta $ itself must vanish at genus two for odd values of $n$. Examples up to $n=4$ can be found in appendix E.

\vspace{\mylength}
\section{Conclusion and outlook}
\vspace{\myotherlength}

The descent procedure described in this work re-organizes cyclic products of Szeg\"o kernels such that their dependences on the points $z_1,\cdots,z_n$ and on the even spin structure $\delta$ are completely disentangled. The key results apply to Riemann surfaces of arbitrary genus which find an increasingly universal appearance in different areas of theoretical physics and mathematics.

\sm

First, the function space identified through the $\delta$-independent building blocks of this work plays a crucial role in the explicit evaluation of multi-particle string amplitudes beyond genus one, as well as in bootstrap approaches to the construction of such amplitudes \cite{Berkovits:2022ivl}. Their intimate connection with the higher-genus polylogarithms of \cite{DHoker:2023vax} will strengthen the symbiosis between algebraic geometry, string perturbation theory and particle physics, stimulating for instance the application to higher-genus surfaces in Feynman integrals \cite{Huang:2013kh, Georgoudis:2015hca, Doran:2023yzu, Marzucca:2023gto}.
 
 \sm
  
Second, the descent procedure introduced here 
sheds further light on the cohomology structure of chiral blocks investigated in \cite{DHoker:2007csw} as the cyclic products $C_\delta(\bz)$ are automatically part of the chiral-block structure \cite{DHoker:1988pdl, DHoker:1989cxq}. Accordingly, the field-theory limit will translate our simplifications of $C_\delta(\bz)$ into new double-copy representations of multi-loop amplitudes in supergravity theories, expressed via bilinears of gauge-theory building blocks \cite{Bern:2019prr, Adamo:2022dcm}.

\sm 

Among the numerous directions of follow-up research, the results of this work suggest future investigations of the following problems:
(a) to simplify the spin-structure sums in the proposal of \cite{Geyer:2021oox} for the genus-three four-point superstring amplitude by means of the four-point results in appendix~B; (b) to explore generalizations of the structure obtained in (\ref{majsimp}) to higher genus, i.e.\ whether the $\delta$-dependence of $C^{I_1 \cdots I_n} _\delta$ at arbitrary $n$ and fixed $h\geq 3$ can still be reduced to \textit{finitely} many tensor products of lower-rank tensors; (c) to explicitly compute the modular tensors $C_\delta^{I_1\ldots I_n}$ of $\Gamma_h(2)$, starting with $C_\delta^{IJK}$ at genus $h=3$ and the components of $C_\delta^{IJKL}$ at  $h=2$ beyond the symmetrized ones in appendix~E.

\sm

{\it Acknowledgements}: The research of ED is supported in part by NSF grant PHY-22-09700. The research of MH is supported in part by the European Research Council under ERC-STG-804286 UNISCAMP and in part by the Knut and Alice Wallenberg Foundation under grant KAW 2018.0116. The research of OS is supported by the European Research Council under ERC-STG-804286 UNISCAMP.

\section{Appendix A. Theta functions and the Szeg\"o kernel}
\vspace{\myotherlength}

In this appendix we define the Riemann $\tet$-function, the prime form, and the Szeg\"o kernel for arbitrary  genus~$h$. For given characteristics $\kappa = [\kappa', \kappa'']$ with $\kappa', \kappa'' \in \mathbb C^h$, the Riemann $\tet$-function of rank $h$ is defined by,
\beq
\vartheta[\kappa](\zeta | \Omega) = \! \sum_{ n \in \mathbb Z^h } \!  
e^{ \pi i  (n{+}\kappa' )^t \Omega (n{+} \kappa')  + 2\pi i (n{+} \kappa' )^t (\zeta {+} \kappa'') } \, ,
\label{appth.01}
\eeq
where $\zeta \in \mathbb C^h$. Henceforth, the dependence on the period matrix $\Omega$  will be suppressed. By specializing $\kappa$ to odd half-characteristics
or spin structures $\nu = [\nu', \nu'']$ with entries $\in \{ 0,\frac{1}{2} \}$ and odd values of $4 \nu' \cdot \nu''$, we 
obtain the prime form \cite{Fay:1973},
\beq
E(x,y) = \frac{ \vartheta[\nu]\big( \int^x_y \omega \big)}{h_\nu(x) h_\nu(y) } \, .
\label{appth.02}
\eeq
This definition is independent of the choice of the odd spin structure $\nu$ by  virtue of the holomorphic $(\frac{1}{2},0)$-forms $h_\nu(x) $ subject  to $h_\nu(x)^2 = \sum_{I=1}^h \omega_I(x) \frac{\partial}{\partial \zeta_I} \vartheta[\nu](0)$.
Finally the Szeg\"o kernel for even spin structure $\delta = [\delta', \delta'']$ (with entries $\in \{ 0,\frac{1}{2} \}$ and even $4 \delta' \cdot \delta''$) is given~by
\beq
S_\delta(x,y) = \frac{ \vartheta[\delta]\big( \int^x_y \omega \big)}{ \vartheta[\delta](0) E(x,y) }
\label{appth.02}
\eeq
at generic moduli.

\vspace{\mylength}
\section{Appendix B. Descent for $n=4$}
\vspace{\myotherlength}

In this appendix, we shall provide the results of the descent procedure at $n=4$ 
to illustrate the case of general $n$ presented in the main text.
The recursive relations produced by the descent procedure are as follows, 
%
\begin{align}
C_\delta (1,2,3,4)  & = 
\om_I(1) C_\delta^I  (2,3,4)    \label{4ptcdel} \\
&\quad
- \big ( \p_1 \cG(1,2) - \p_1 \cG(1,4) \big ) C_\delta (2,3,4) \, ,
\no \\
C_\delta ^I(2,3,4) & =  \om_J(2) C^{IJ} _\delta(3,4) + f^I{}_{\!J}(2,3) C^J_\delta(3,4)  \notag \\
&\quad
- \big ( \p_2 \cG^I(2,3) - \p_2 \cG^I(2,4) \big ) C_\delta(3,4) \, ,
\no \\
C^{IJ}_\delta(3,4) & =  \om_K(3) C^{IJK}_\delta(4) + f^J{}_{\!K} (3,4) C^{IK}_\delta(4)  \notag \\
&\quad
  + f^{JI}{}_{\!K} (3,4) C^{KL} _\delta \om_L(4) + \p_3 \p_4 \cG^{JI} (3,4) \, ,
\no \\
C^{IJK} _\delta (4) & = 
\om_L(4) C^{IJKL} _\delta  \no \\ 
&\quad
- \p_4 \Phi^I{}_{\!L}(4) C_\delta ^{JKL} 
+ \p_4 \Phi^K{}_{\!L}(4) C^{IJL}_\delta 
\no \\ 
&\quad
- C_\delta ^{JL}  \p_4 \Phi^{IK}{}_{\!L} (4)
 - C_\delta ^{JL}   \p_4 \Phi^{KI}{}_{\!L} (4)  \no \\ 
&\quad
 + C_\delta ^{KL}  \p_4 \Phi^{IJ}{}_{\!L} (4) 
 + C^{IL} _\delta  \p_4 \Phi^{KJ}{}_{\!L} (4)  \, .
 \notag
\end{align}
The differential relations analogous to (\ref{1.three.a}), (\ref{1.three.b}) and (\ref{antiholo3}) can be found in appendix~C. 
Analogously to (\ref{int.three}), we can directly express $C^{IJKL}_\delta$ as the following integral,
\begin{align}
\label{int.four}
C_\delta ^{IJKL} & =   
\int _\Sigma d^2 w \, \bar \om ^I(w)  \int _\Sigma d^2 x \, \bar \om ^J(x) \int _\Sigma d^2 y \, \bar \om ^K(y)
 \notag 
\\ 
&\quad \times  \int _\Sigma d^2 z \, \bar \om ^L(z) \, C_\delta(w,x,y,z)
\end{align}
and expose the following reflection and cyclic symmetry,
\bea
C_\delta^{JKLI} = C_\delta^{JILK} = C_\delta ^{IJKL} \, .
\eea
Successive elimination of $C^I_\delta(2,3,4)$, $C^{IJ}(3,4)$ and $C^{IJK}_\delta(4)$ in (\ref{4ptcdel}) along with the elimination of $C_\delta (3,4)$ and $C_\delta(2,3,4)$ using (\ref{1.two}) and (\ref{fat3pt}), respectively, proves that $C_\delta(1,2,3,4)$ is a linear combination of the modular tensors $C^{IJKL}_\delta$, $C^{IJK}_\delta$, and $C^{IJ}_\delta$ of $\Gamma_h(2)$, with coefficients that are independent of the spin structure $\delta$ and contain all the dependence on the points $z_1, \cdots, z_4$. 

\section{Appendix C. Differential relations for the descent procedure}
\vspace{\myotherlength}


The decompositions (\ref{alln.00}) to (\ref{alln.02}) of the $n$-point cyclic product $C_\delta(\bz)$ at arbitrary genus can be derived from the following differential relations,
\begin{widetext}

$\bullet$ For  $i=1,\cdots, n$, identifying $n{+}1 \equiv 1$, and $0 \equiv n$, 
\bea
\bar \p_i C_\delta(1, \cdots, n)  & =  & 
\pi \big ( \delta (i,i{+}1) - \delta (i,i{-}1) \big ) 
C_\delta (1, \cdots, \hat i, \cdots , n)  \, .
\quad
\label{df.1}
\eea

$\bullet$ For  $j=2,\cdots, n{-}1$ and   $ j{+}1 \leq i \leq n$,
\bea
\bar \p_i C_\delta^{I_1 \cdots I_{j-1}}  (j, \cdots, n)  
= \pi \big ( \delta (i,i{+}1) - \delta (i,i{-}1) \big ) 
C_\delta ^{I_1 \cdots I_{j-1}}  (j, \cdots, \hat i, \cdots , n)  \, .
\label{df.2}
\eea

$\bullet$ For $j=2,\cdots, n{-}1$,
\bea
\bar \p_j C_\delta^{I_1 \cdots I_{j-1}}  (j, \cdots, n)   
& = & \pi  \delta (j,j{+}1) C_\delta ^{I_1 \cdots I_{j-1}}  (j{+}1,  \cdots , n) 
- \pi \bar \om ^{I_{j-1}} (j) C_\delta ^{I_1 \cdots I_{j-2}} (j, \cdots , n)  \, ,
\no \\
 \bar \p_n C^{I_1 \cdots I_{j-1}} _\delta (j, \cdots, n)  
&=& - \pi \delta (n,n{-}1) C^{I_1 \cdots I_{j-1}} _\delta (j, \cdots , n{-}1) 
+ \pi \bar \om ^{I_1} (n) C_\delta ^{I_2 \cdots I_{j-1}}  (j, \cdots, n) \, ,
\label{df.3}
\eea
\end{widetext}
which specialize as follows at $n=4$ points:
\bea
\bar \p_1 C_\delta(1,2,3,4) & = & \pi \big ( \delta(1,2) - \delta (1,4) \big ) C_\delta(2,3,4) \, ,
 \\
\bar \p_2 C^I_\delta(2,3,4) & = &\pi \delta(2,3) C^I_\delta(3,4) - \pi \bar \om^I(2) C_\delta(2,3,4) \, ,
\no \\
\bar \p_3 C^{IJ} _\delta (3,4) & = & \pi \delta (3,4) C^{IJ}_\delta(4) - \pi C^I_\delta (3,4) \bar \om^J(3) \, ,
\no \\
\bar \p_4 C^{IJK}_\delta(4) & = & \pi \bar \om^I(4) C^{JK}_\delta(4) - \pi \bar \om^K(4) C^{IJ} _\delta(4) \, .
\no
\eea

\vspace{\mylength}
\section{Appendix D. Examples of the $C^{I_1\cdots I_n}_\delta$-tensors at genus one}
\vspace{\myotherlength}

In this appendix, we shall display the $(n\leq 8)$-point genus-one examples of the $\Gamma_{h=1}(2)$ tensors $C_\delta^{11\cdots 1}$ defined by (\ref{int.n}). A first dependence on the modulus $\tau = \Omega_{11}$ occurs via holomorphic Eisenstein series ${\rm G}_w(\tau)$ which furnish modular forms of weight $(w,0)$ if ${w\geq 4}$ as well as the almost holomorphic modular form of weight $(2,0)$,
\bea
\hat {\rm G}_2(\tau) = {\rm G}_2(\tau)  - \tfrac{ \pi }{\Im \tau} \, ,
\eea
see for instance \cite{123modular, DHoker:2022dxx} and references therein.
Moreover, the entire $\delta$-dependence of the cycles $C_\delta(\bz)$ at genus one and arbitrary multiplicity is captured by the Weierstrass $\wp$ function evaluated at half-periods, 
\begin{align}
e_\delta(\tau) &\in \{ \wp\big(\tfrac{1}{2}|\tau\big), \wp\big(\tfrac{\tau}{2} |\tau\big), \wp\big(\tfrac{1+\tau}{2} |\tau\big) \}  \, .
 \end{align}
The building blocks ${\rm G}_w,e_\delta$ admit the following compact expressions for
the genus-one instances $C_\delta^{11\cdots 1}$ of the modular tensors (\ref{int.n}) up to
and including $n=8$ points,
\begin{align}
C_\delta^{11} &= e_\delta {+} \hat {\rm G}_2 \, , &\ \,
C_\delta^{111111} &=  6e_\delta {\rm G}_4 {+} 21 {\rm G}_6  \, , \\
C_\delta^{1111} &= e_\delta^2  {-} 5 {\rm G}_4  \, ,
&\ \, C_\delta^{11111111} &=  3 e_\delta^2 {\rm G}_4 {+} 15 e_\delta {\rm G}_6 {+} 15 {\rm G}_8 \, .
\notag
\end{align}
More general even multiplicities $n\geq 4$ lead to degree-two polynomials in $e_\delta$,

\vspace{-0.42cm}
\bea
C_\delta^{\overbrace{11\cdots 1}^n} = e_\delta^2\,  \varphi_{n-4}
+ e_\delta\, \psi_{n-2} +    \rho_{n}  \, ,
\label{g1npt}
\eea
with holomorphic modular forms $\varphi_w,\psi_w,\rho_w$ of $SL(2,\mathbb Z)$ of weight $(w,0)$,
whereas $C_\delta^{11\cdots 1}$ at odd $n$ vanish.

\vspace{\mylength}
\section{Appendix E. Examples of the $C^{I_1\cdots I_n}_\delta$-tensors at genus two}
\vspace{\myotherlength}

This appendix is dedicated to the genus-two cases of the modular tensors $C^{I_1\cdots I_n}_\delta$  in (\ref{int.n}) whose spin-structure dependence at arbitrary $n$ is expressible via degree-two polynomials in the tensor $\mL_\delta^{IJ}$ given by (\ref{defelldelta}) \cite{DHoker:2022xxg}. The two-point results (\ref{rk2c}) and (\ref{Cph2}) lead to the representation
\begin{align}
C_\delta^{IJ} &= \mL_\delta^{IJ} - \tfrac{i\pi}{5} \partial^{IJ} \ln \Psi_{10} - \pi Y^{IJ}
\end{align}
linear in $\mL_\delta^{IJ}$. The next non-vanishing genus-two case of $C^{I_1\cdots I_n}_\delta$ occurs at $n=4$ points, where the results in appendix H of \cite{DHoker:2022xxg} identify the totally symmetrized part,
\begin{align}
&C_\delta^{(IJKL)} = \mL_\delta^{(IJ}  \mL_\delta^{KL)} \label{symm4pt} \\
&\quad
- \tfrac{1}{2} \mM_1^{IJKLMN} \mL_\delta^{PQ} \varepsilon_{MP}\varepsilon_{NQ} -\tfrac{1}{2} \mM_2^{IJKL}
\, , \notag
\end{align}
involving the antisymmetric $\varepsilon_{IJ}$-tensor with $ \varepsilon_{12} = 1$.
The totally symmetric $Sp(4,\mathbb Z)$ modular tensors $\mM_i$ are obtained by translating the $SL(2,\mathbb C)$-tensors ${\bM}_i$ of \cite{DHoker:2022xxg} into the language of $\tet$-functions,
\begin{align}
\mM_1^{I_1  \cdots I_6} &= \frac{ 4 \pi^{-4} }{ \sqrt{ \Psi_{10}} } 
\partial^{(I_1} \tet[\nu_1](0)
 \partial^{I_2} \tet[\nu_2](0)
 \cdots
  \partial^{I_6)} \tet[\nu_6](0)  \, ,\notag \\
  \mM_2^{I_1 \cdots I_4}  & = 
\tfrac{1}{2}  \mM_1 ^{J_1 \cdots J_4( I_1 I_2 }  \mM_1^{I_3I_4) K_1 \cdots K_4} 
\ep _{J_1 K_1} \! \! \cdots\! \ep _{J_4 K_4}  \, ,  
\end{align}
where $\nu_1,\cdots,\nu_6$ denote the six distinct odd spin structures at genus two;
we define $\sqrt{ \Psi_{10}} = \prod_{\delta \, {\rm even}} \tet[\delta](0)$; and indices inside parentheses are totally symmetrized.

\vspace{\mylength}


\vskip -0.3cm

\begin{thebibliography}{99}
\vspace{-0.2cm}

\bibitem{Mumford}
D. Mumford, 
Progress in Math. {\bf 28} (1983), Birkh\"auser.

\bibitem{Fay:1973}
J.~D.~Fay, Lecture Notes in Math. {\bf 352} (1973).

\bibitem{Tsuchiya:1988va}
A.~G.~Tsuchiya,
Phys. Rev. D \textbf{39} (1989), 1626.

\bibitem{Stieberger:2002wk}
S.~Stieberger and T.~R.~Taylor,
Nucl. Phys. B \textbf{648} (2003), 3-34
[arXiv:hep-th/0209064].

\bibitem{Bianchi:2006nf}
M.~Bianchi and A.~V.~Santini,
JHEP \textbf{12} (2006), 010
[arXiv:hep-th/0607224].

\bibitem{Dolan:2007eh}
L.~Dolan and P.~Goddard, 
Commun. Math. Phys. {\bf 285} (2009) 219-264,
 arXiv.:0710.3743.

\bibitem{Tsuchiya:2012nf}
A.~G.~Tsuchiya,
arXiv:1209.6117.

\bibitem{Broedel:2014vla}
J.~Broedel, C.~R.~Mafra, N.~Matthes and O.~Schlotterer,
JHEP \textbf{07} (2015), 112,
arXiv:1412.5535.

\bibitem{Berg:2016wux}
M.~Berg, I.~Buchberger and O.~Schlotterer,
JHEP \textbf{04} (2017), 163,
arXiv:1603.05262.

\bibitem{Tsuchiya:2017joo}
A.~G.~Tsuchiya,
arXiv:1710.00206.


\bibitem{DHoker:2022xxg}
E.~D'Hoker, M.~Hidding and O.~Schlotterer,
JHEP \textbf{05} (2023), 073,
arXiv:2211.09069.

\bibitem{DHoker:2001jaf}
E.~D'Hoker and D.~H.~Phong,
Nucl. Phys. B \textbf{639} (2002), 129-181,
[arXiv:hep-th/0111040].

\bibitem{DHoker:2005vch}
E.~D'Hoker and D.~H.~Phong,
Nucl. Phys. B \textbf{715} (2005), 3-90,
[arXiv:hep-th/0501197].

\bibitem{DHoker:2021kks}
E.~D'Hoker and O.~Schlotterer,
JHEP \textbf{12} (2021), 063,
arXiv:2108.01104.

\bibitem{Tsuchiya:2022lqv}
A.~G.~Tsuchiya,
Nucl. Phys. B \textbf{997} (2023), 116383,
arXiv:2209.14633.




\bibitem{DHoker:2023vax}
E.~D'Hoker, M.~Hidding and O.~Schlotterer,
arXiv:2306.08644.

\bibitem{Mafra:2022wml}
C.~R.~Mafra and O.~Schlotterer,
Phys. Rept. \textbf{1020} (2023), 1-162.


\bibitem{Falt}
G.~Faltings, Ann. Math. {\bf 119} (1984), 387.

\bibitem{DHoker:2017pvk}
E.~D'Hoker, M.~B.~Green and B.~Pioline,
Commun. Math. Phys. \textbf{366} (2019) no.3, 927-979,
arXiv:1712.06135.

\bibitem{DHoker:2020uid}
E.~D'Hoker and O.~Schlotterer,
Commun. Num. Theor. Phys. \textbf{16} (2022) no.1, 35-74,
arXiv:2010.00924.

\bibitem{Kawazumi:lecture}
N.~Kawazumi, 
\href{http://www.ms.u-tokyo.ac.jp/~kawazumi/OIST1610_v1.pdf}{Lecture at MCM2016} , OIST (2016).

\bibitem{Kawazumi:paper}
N.~Kawazumi, arXiv:2210.00532 [math.GT].


\bibitem{Enriquez:2011}
B.~Enriquez, Adv. Math. {\bf 252} (2014), 204-226, arXiv:1112.0864 [math.GT].

\bibitem{Enriquez:2021}
B.~Enriquez and F.~Zerbini,
arXiv:2110.09341 [math.AG].

\bibitem{Enriquez:2022}
B.~Enriquez and F.~Zerbini,
arXiv:2212.03119 [math.AG].



\bibitem{Verlinde:1986kw}
E.~P.~Verlinde and H.~L.~Verlinde,
Nucl. Phys. B \textbf{288} (1987), 357.

\bibitem{Verlinde:1987sd}
E.~P.~Verlinde and H.~L.~Verlinde,
Phys. Lett. B \textbf{192} (1987), 95-102.

\bibitem{DHoker:1988pdl}
E.~D'Hoker and D.~H.~Phong,
Rev. Mod. Phys. \textbf{60} (1988), 917.



\bibitem{DHoker:2014oxd}
E.~D'Hoker, M.~B.~Green, B.~Pioline and R.~Russo,
JHEP \textbf{01} (2015), 031,
arXiv:1405.6226.

\bibitem{Berkovits:2022ivl}
N.~Berkovits, E.~D'Hoker, M.~B.~Green, H.~Johansson and O.~Schlotterer,
arXiv:2203.09099.

\bibitem{Huang:2013kh}
R.~Huang and Y.~Zhang,
JHEP \textbf{04} (2013), 080,
arXiv:1302.1023.

\bibitem{Georgoudis:2015hca}
A.~Georgoudis and Y.~Zhang,
JHEP \textbf{12} (2015), 086,
arXiv:1507.06310.

\bibitem{Doran:2023yzu}
C.~F.~Doran, A.~Harder, E.~Pichon-Pharabod and P.~Vanhove,
arXiv:2302.14840.

\bibitem{Marzucca:2023gto}
R.~Marzucca, A.~J.~McLeod, B.~Page, S.~P\"ogel and S.~Weinzierl,
Phys. Rev. D \textbf{109} (2024) no.3, L031901,
arXiv:2307.11497.

\bibitem{DHoker:2007csw}
E.~D'Hoker and D.~H.~Phong,
Nucl. Phys. B \textbf{804}, 421-506 (2008),
arXiv:0711.4314.


\bibitem{DHoker:1989cxq}
E.~D'Hoker and D.~H.~Phong,
Commun. Math. Phys. \textbf{125} (1989), 469.

\bibitem{Bern:2019prr}
Z.~Bern, J.~J.~Carrasco, M.~Chiodaroli, H.~Johansson and R.~Roiban,
arXiv:1909.01358.

\bibitem{Adamo:2022dcm}
T.~Adamo, J.~J.~M.~Carrasco, M.~Carrillo-Gonz\'alez, M.~Chiodaroli, H.~Elvang, H.~Johansson, D.~O'Connell, R.~Roiban and O.~Schlotterer,
arXiv:2204.06547.

\bibitem{Geyer:2021oox}
Y.~Geyer, R.~Monteiro and R.~Stark-Much\~ao,
Phys. Rev. Lett. \textbf{127} (2021) no.21, 211603,
arXiv:2106.03968.

\bibitem{123modular}
J.~H.~Bruinier, G.~van der Geer, G.~Harder and D.~Zagier,
``The 1-2-3 of Modular Forms'',
(2008), Springer.

\bibitem{DHoker:2022dxx}
E.~D'Hoker and J.~Kaidi,
arXiv:2208.07242.

\end{thebibliography}
\end{document}